\documentclass[showpacs,preprintnumbers]{revtex4} 

\usepackage{graphicx}
\usepackage{dcolumn}
\usepackage{epsf} 
\usepackage{latexsym} 

\def\beq{\begin{equation}}
\def\eeq{\end{equation}}
\def\rmd{{ d}}
\def\fl{{}}

\input prepictex 
\input pictex 
\input postpictex
\def\mathput#1{\relax \ifmmode \displaystyle #1\else $\displaystyle #1$\fi}

\def\pmb#1{\setbox0=\hbox{$#1$}%
  \kern-.025em\copy0\kern-\wd0
  \kern.05em\copy0\kern-\wd0
  \kern-.025em\raise.0433em\box0}
 
\begin{document}

\title{Limitations of Radar Coordinates}

\author{D. Bini}
\affiliation{Istituto per le Applicazioni del Calcolo ``M. Picone'', CNR, I-00161 Rome, Italy and\\ 
International Center for Relativistic Astrophysics - I.C.R.A.\\
University of Rome ``La Sapienza'', I-00185 Rome, Italy\\
 INFN Sezione di Firenze, via G. Sansone, 1, I-50019 Sesto Fiorentino (FI), Italy}

\author{L. Lusanna}
\affiliation{ INFN Sezione di Firenze, via G. Sansone, 1, I-50019 Sesto Fiorentino (FI), Italy
}

\author{B. Mashhoon}
\affiliation{Department of Physics and Astronomy,
University of Missouri-Columbia, Columbia,
Missouri 65211, USA}

\date{\today}

\begin{abstract}
The construction of a radar coordinate system about the world line of an observer is discussed. Radar coordinates for a hyperbolic observer as well as a uniformly rotating observer are described in detail. The utility of the notion of radar distance and the admissibility of radar coordinates are investigated. Our results provide a critical assessment of the physical significance of radar coordinates.
\end{abstract}


\maketitle

\section{Introduction}

A physical observer's most basic measurements involve the determination of temporal and spatial intervals.
The fundamental nongravitational laws of physics have been formulated with respect to ideal inertial observers; therefore, let us first discuss the spacetime measurements of inertial observers. In an inertial frame of reference with Cartesian coordinates $x^\mu=(t,x,y,z)$, the fundamental observers are the ideal inertial observers that are at rest in the global reference system. Such observers have access to ideal clocks and measuring rods.
The clocks are assumed to be synchronized; for instance, two adjacent clocks at rest can be synchronized by a fundamental observer and then one of the clocks can be adiabatically transported to another location. The transport can be so slow as to have no practical impact on the synchronization of clocks. Similarly, lengths are determined in general by placing infinitesimal measuring rods together. The measurements of uniformly moving inertial observers are related to those of the fundamental observers by Lorentz invariance, which leads to the phenomena of length contraction and time dilation \cite{1}. To apply these elementary ideas in realistic situations involving accelerated systems and gravitational fields, certain generalizations must be considered.  
In fact, all actual physical observers are more or less noninertial. The standard generalization of the Cartesian inertial coordinates to the case of observers in accelerated systems and gravitational fields involves the introduction of Fermi coordinates \cite{2}. These constitute a geodesic coordinate system $X^\mu=(T,X,Y,Z)$ that is established along the world line of a reference observer that carries an orthonormal tetrad frame along its path.
At each event on the path characterized by the observer's proper time $\tau$, a hypersurface normal to the world line is constructed by all spacelike geodesics normal to the world line. On these hypersurfaces of simultaneity, lengths away from the world line are defined using the proper length of the spacelike geodesic. This is based on the summation of infinitesimal meter sticks placed together along a locally straight line. We note that the underlying assumption regarding these measurements of a noninertial observer in the standard theory of relativity is the hypothesis of locality. That is, the noninertial observer is pointwise equivalent to an otherwise identical momentarily comoving inertial observer \cite{3}.

In an inertial reference frame, the  above construction along a straight world line with a parallel-propagated frame generates the entire Minkowski spacetime. However, for accelerated systems and gravitational fields there are 
well-known limitations that are related to the existence of invariant acceleration and curvature lengths \cite{3,4}.

Since the early days of relativity theory, an alternative method of synchronization of distant clocks --- as well as the measurement of the distance between them --- has been discussed based on the transmission and reception of light signals.
Specifically, imagine an arbitrary observer P on a reference world line and suppose that at its proper time $\tau_1$ it transmits a light signal to a (moving) observer Q that immediately and without delay transmits a light signal right back to P as in figure 1. The second signal is received by P at its proper time $\tau_2$. We suppose that the clock carried by Q at the event of signal interchange is simultaneous with the event along the world line of the reference observer P at its proper time $\eta=(\tau_1+\tau_2)/2$. In this way the clocks of P and Q are synchronized. That is $\eta-\tau_1= \tau_2-\eta$, so that the time $(\eta-\tau_1)$ that it takes for the first signal to reach Q is equal to the time $(\tau_2-\eta)$ that it takes for the second signal to reach P. The corresponding distance from P to Q at the instant of simultaneity is then $\rho=c(\tau_2-\tau_1)/2$. The limiting case of $\tau_1 = \tau_2$ refers to the location of observer P, where $\rho = 0$ and $\eta = \tau$ is the proper time of the reference observer.
 It is well known that for inertial observers in Minkowski spacetime, these operational procedures are equivalent to  the standard approach discussed above. Moreover, this method can be employed in accelerated systems and gravitational fields for observers
whose world lines are infinitesimally close to each other \cite{5}; the results are then identical to those obtained by using Fermi coordinates. On the other hand, it has been suggested that this \lq\lq radar" approach may also be useful for general observers \cite{6}. The synchronization of distant clocks by light signals may lead to a foliation of spacetime by spacelike hypersurfaces of fixed $\eta$. Indeed  in view of the well-known limitations of Fermi coordinates in terms of their admissibility, some authors have recently suggested that radar coordinates may be preferable \cite{7,8}. The purpose of this paper is to demonstrate that this is not the case by pointing out the limitations of radar coordinates.
We confine our main discussion to noninertial observers in Minkowski spacetime for the sake of simplicity. In section 2 we define radar coordinates and discuss critically the concept of radar distance. Hyperbolic and uniformly rotating observers are considered in sections 3 and 4, and the limited domains of admissibility of their radar coordinates are demonstrated. Section 5 contains a brief discussion of our results.
Fermi coordinates for a uniformly rotating observer and the standard admissibility conditions are discussed in appendices A and B, respectively.

\section{Radar coordinates}

Imagine an observer P following a world line $\bar x{}^\mu(\tau)$ in a global inertial coordinate system. Henceforth we use units such that $c=1$.  The radar time and distance that P assigns to observer Q at $x^\mu=(t,x,y,z)$ are $\eta$ and $\rho$, respectively. Thus in figure 1, $\tau_1=\eta -\rho$, $\tau_2=\eta+\rho$ and the equations for the null rays in figure 1 imply that 
\begin{eqnarray}
\label{eq:1}\fl\quad
t- \bar t (\eta-\rho )&= \sqrt{[x-\bar x(\eta-\rho )]^2+[y-\bar y(\eta-\rho )]^2+[z-\bar z(\eta-\rho )]^2} ,\\
\label{eq:2}\fl\quad
t- \bar t (\eta+\rho )&= -\sqrt{[x-\bar x (\eta+\rho )]^2+[y-\bar y(\eta+\rho ) ]^2+[z-\bar z(\eta+\rho )]^2} .
\end{eqnarray}
These equations basically define $\eta$ and $\rho$ in terms of $t$, $x$, $y$ and $z$.
We define the radar coordinates of Q to be $(\eta, \rho, \theta, \phi)$, where $(\theta, \phi)$ are the standard angular coordinates that uniquely identify the direction from P to Q; therefore,  $-\infty <\eta <\infty$,
$0\le \rho < \infty$, $0\le \theta \le \pi$, $0\le \phi < 2\pi$. The radar coordinate system  is thus a \lq\lq spherical" polar coordinate system  that is set up such that P is fixed at the origin of the spatial coordinates.
As usual, the proper domain of definition of these polar coordinates excludes their associated $z-$axis.

\begin{figure}[h]
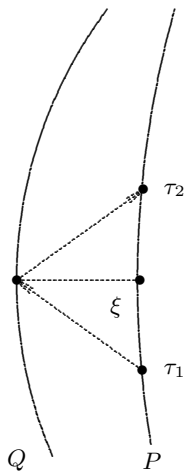

$$ \vbox{
\beginpicture
  \setcoordinatesystem units <0.6cm,0.6cm> point at 0 0 

\put {\mathput{\phantom{.}}}         at 2.8 3

\put {\mathput{\bullet}}            at 2.8 -1 
\put {\mathput{\bullet}}            at 2.73 -3 
\put {\mathput{\bullet}}            at 2.78 -5 
\put {\mathput{\bullet}}            at 0 -3
\put {\mathput{P}}          at 3 -7
\put {\mathput{Q}}          at 0 -7
\put {\mathput{\tau_2}}          at 3.5 -1
\put {\mathput{\tau_1}}          at 3.5 -5
\put {\mathput{\xi}}          at 2.2 -3.6  

\setdashes <1pt>
\arrow <.25cm> [.2,.3]    from  0 -3  to 2.8 -1
\arrow <.25cm> [.2,.3]    from  2.78 -5  to 0 -3
\arrow <.0cm> [0,0]    from  0 -3 to 2.73 -3

\setsolid
\circulararc 60 degrees from 2 3 center at 10 -3
\circulararc 25 degrees from 3.5 3 center at 25 -3

\endpicture}$$
\caption{Schematic representation of the interchange of light signals by observers P and Q.}
\end{figure}

Imagine, for instance, that P is an inertial observer moving with constant speed $v$ along the $x-$direction
\beq
\bar t = \gamma \tau, \quad \bar x= \gamma v \tau, \quad \bar y=0, \quad \bar z=0,
\end{equation}
where $\gamma=1/\sqrt{1-v^2}$.
Then (\ref{eq:1}) and (\ref{eq:2}) reduce to
\begin{eqnarray}
\label{eq:4} \fl\qquad
&& t-\gamma(\eta-\rho)=\sqrt{[x-\gamma v(\eta-\rho)]^2+
y^2+z^2},\\
\label{eq:5}\fl\qquad
&&  t-\gamma(\eta+\rho)=-\sqrt{[x-\gamma v(\eta+\rho)]^2+
y^2+z^2}.  
\end{eqnarray}
After squaring (\ref{eq:4}) and (\ref{eq:5}) and then adding and subtracting the resulting equations we find the transformation between $(t,x,y,z)$ and $(\eta, \rho, \theta, \phi)$,
\begin{eqnarray}
\label{eq:6}
t&=&\gamma (\eta +v\rho \sin \theta \cos \phi), \nonumber \\
x&=&\gamma (v\eta +\rho \sin \theta \cos \phi), \nonumber \\
y&=& \rho \sin \theta \sin \phi, \nonumber \\
z&=&\rho \cos \theta,
\end{eqnarray}
such that for $v=0$ we recover the standard inertial coordinates where the spatial coordinates are expressed using a spherical polar coordinate system.
This result has a simple physical interpretation. First consider the Lorentz boost $(t,x,y,z) \to (t',x',y',z')$, where P is at the spatial origin of the new coordinates
\beq
\label{eq:7}
t=\gamma(t'+vx'),\quad x=\gamma (x'+vt'), \quad y=y',\quad z=z'.
\end{equation}
Next, consider the transformation to spherical polar coordinates in the boosted frame, i.e. $(t',x',y',z')\to (\eta, \rho, \theta, \phi)$,
\beq
\label{eq:8}
t'=\eta,\quad x'=\rho \sin \theta \cos \phi, \quad y'=\rho \sin \theta \sin \phi, \quad z'=\rho \cos \theta .
\end{equation}
Substituting (\ref{eq:8}) in (\ref{eq:7}) results in (\ref{eq:6}). 
It proves useful to define the straight line segment $\xi^\mu=x^\mu-\bar x{}^{\mu}(\eta) $ that connects simultaneous events on the world lines of P and Q
as in figure 1. The components of this vector with respect to the boosted frame are given by $\xi'{}^{\mu}=(0,x',y',z')$; therefore, the angular coordinates $(\theta, \phi)$ can be defined by the direction of the vector $\xi$ with respect to the inertial frame comoving with the observer.
It is thus clear 
that for an inertial observer radar coordinates can in principle cover the whole Minkowski spacetime and $\rmd s^2=\eta_{\alpha\beta}\rmd x^\alpha \rmd x^\beta$ is then given by
\beq
\label{eq:9}
\rmd s^2=-\rmd \eta^2+\rmd \rho^2+ \rho^2 (\rmd \theta^2 + \sin^2 \theta \rmd \phi^2).
\end{equation}

To generalize our approach to include accelerated observers in gravitational fields, it is advantageous to introduce the natural tetrad frame of the observer. For instance, let $\lambda^\mu{}_{(\alpha )}$ be the natural tetrad frame of the inertial observer P:
\begin{eqnarray}
\lambda^\mu{}_{(0)}&= \gamma (1,v,0,0), \qquad \lambda^\mu{}_{(1)}&= \gamma (v,1,0,0), \nonumber \\
\lambda^\mu{}_{(2)}&= \gamma (0,0,1,0), \qquad \lambda^\mu{}_{(3)}&= \gamma (0,0,0,1).
\end{eqnarray}
The components of the vector $\xi$ with respect to this frame are given by $\xi_{(\alpha )}=\xi_\mu \lambda^\mu{}_{(\alpha)}$, i.e. $\xi_{(0)}=0$,
$\xi_{(1)}=\rho \sin \theta \cos \phi$, $\xi_{(2)}=\rho \sin \theta \sin \phi$ and $\xi_{(3)}=\rho \cos \theta $. Thus the spherical polar coordinates 
$(\rho , \theta, \phi)$ completely characterize the spatial vector $\xi$ with respect to the natural tetrad frame of the observer. 
Extending this analysis to noninertial observers provides a definite method of choosing the angular coordinates $\theta$ and $\phi$. To solve this problem in general, we introduce a natural tetrad frame for the reference observer P. At each instant of proper time $\tau$, P is at the spatial origin of its local orthonormal triad. The polar angles are then chosen with respect to this local triad, that is,  at each instant $\tau$, $(\rho , \theta, \phi)$ are the spherical polar coordinates of $\xi$ in an infinitesimal spatial neighborhood of P.  

Instead of a background global inertial system, as in the present paper, consider now the motion of the observer P in general curvilinear coordinates.
The light signals will follow null geodesics in this case and the line segment $\xi$ connecting simultaneous events on P and Q will no longer be a vector; in fact, it must be replaced by a (spacelike) geodesic. Let $\hat \xi{}^\mu$ be the {\it unit} tangent vector to this geodesic at P; then, the radar angles $\theta$ and $\phi$ will be defined such that they locally characterize $\hat \xi{}^\mu$ in the natural spatial frame at P, i.e.
\begin{eqnarray}
\hat \xi{}_\mu \, \lambda^\mu{}_{(1)}&=& \sqrt{1+\delta^2}\sin \theta \cos \phi\, , \nonumber \\
\hat \xi{}_\mu \, \lambda^\mu{}_{(2)}&=& \sqrt{1+\delta^2}\sin \theta \sin \phi\, , \nonumber \\
\hat \xi{}_\mu \, \lambda^\mu{}_{(3)}&=& \sqrt{1+\delta^2}\cos \theta \, , 
\end{eqnarray}
where $\delta :=\hat \xi{}_\mu \, \lambda^\mu{}_{(0)}$. As Q approaches P, $\delta \to 0$; indeed, in the immediate neighborhood of P, radar coordinates $(\eta , \rho , \theta , \phi)$ essentially reduce to Fermi coordinates $(T,X,Y,Z)$ such that $T=\eta$, $X=\rho \sin \theta \cos \phi$,
$Y=\rho \sin \theta \sin \phi$, $Z=\rho \cos \theta $ for $\rho \ll L$, where $L$ is a characteristic lengthscale. It follows that the hypersurfaces of radar simultaneity are always orthogonal to the world line of P.

The principal advantage of radar coordinates has to do with the nonlocal synchronization of distant clocks as well as a definition of radial distance based on a certain average light travel time.
We now wish to show that the utility of the latter is  quite limited. To see the problem with radar distance,  consider an observer P rotating uniformly on a circle of radius $r$ about an observer Q as in figure 2. If a light signal traverses the radius of the circle in a time $t$, $t=r$, then $\tau_2-\tau_1=2t\sqrt{1-v^2}$, where $v$ is the uniform speed of the observer P.  Thus we have the unusual result that
\beq
\label{eq:10}
\rho=r\sqrt{1-v^2}.
\end{equation}

\begin{figure}[h]
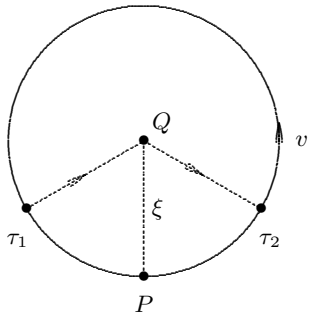

$$ \vbox{
\beginpicture
  \setcoordinatesystem units <0.6cm,0.6cm> point at 0 0 
\circulararc 360 degrees from 3 -3 center at 0 -3

\put {\mathput{\phantom{.}}}         at 2.8 3

\put {\mathput{\bullet}}            at 2.6 -4.5 
\put {\mathput{\bullet}}            at -2.6 -4.5 
\put {\mathput{\bullet}}            at 0 -3
\put {\mathput{\bullet}}            at 0 -6

\put {\mathput{Q}}          at 0.4 -2.6
\put {\mathput{P}}          at 0 -6.6
\put {\mathput{\tau_1}}          at -2.8 -5.2
\put {\mathput{\tau_2}}          at 2.8 -5.2
\put {\mathput{\xi}}          at 0.3 -4.5
\put {\mathput{v}}         at 3.5 -3
\arrow <.25cm> [.2,.3]    from  3 -3  to 3 -2.6

\setdashes <1pt>
\arrow <.25cm> [.2,.3]    from  -2.6 -4.5  to -1.3 -3.75
\arrow <.25cm> [.2,.3]    from  0 -3  to 1.3 -3.75
\plot 0 -3  0 -6 /
\plot  -1.3 -3.75   0 -3  /
\plot  1.3 -3.75   2.6 -4.5 / 
\setsolid

\endpicture}$$
\caption{Schematic representation of the determination of the radar distance from P to Q.}
\end{figure}

The direction of motion of P is always {\it orthogonal} to the direction from P to Q, yet the radar distance to the center 
has a contraction factor of $\sqrt{1-v^2}$ and thus
depends on the speed of motion of P. That is, the radar distance from P to Q is always less than $r$ and approaches zero if the speed of P approaches the speed of light. This problem is absent in Fermi coordinates; in fact, it is shown in appendix A that in terms of Fermi coordinates we get the standard result that the distance from P to Q is $r$. 

In a sufficiently small neighborhood around any regular event in a gravitational field, the spacetime is approximately flat. It then follows from our results that in such a neighborhood the infinitesimal radar distance employed for instance in \cite{5} in fact agrees with the corresponding Fermi distance in that limit; however, for finite separations the Fermi distance is in general different from the radar distance, which must then be employed with great care \cite{9}.

The relationship between the radar and Fermi coordinates can be further clarified as follows: Let $(T, X, Y, Z)$ be the Fermi coordinate system around the world line of observer P. In the immediate neighborhood of P, we introduce the spherical polar coordinates ($\hat \rho$, $\theta$, $\phi$) such that $X = \hat \rho \sin \theta \cos \phi$, $Y =\hat \rho \sin \theta \sin \phi$ and $Z = \hat \rho \cos \theta$. Here $\hat \rho$ is sufficiently small by construction, i.e. $\hat \rho \ll L$, where $L$ is the acceleration length of P. It follows that $(T, \hat \rho , \theta, \phi)$ is an admissible coordinate system and with $T = \eta$ and $\hat \rho = \rho$ coincides with the radar coordinate system around P. Beyond this limiting situation, the two coordinate systems generally differ from each other.
On general grounds, we expect that for $\rho \sim L$ the radar coordinate system may become inadmissible. 
The standard admissibility conditions are discussed in \cite{5}; in any case, we must have $g_{\eta \eta} < 0$ in radar coordinates. 
We will show in the next section that for the hyperbolic observer $(L = 1/g)$, the regime of applicability of the two coordinate systems coincide, whereas this is not the case for the uniformly rotating observer $(L = 1/ \Omega)$ discussed in section 4. A brief treatment of the standard admissibility 
conditions is given in appendix B; for a more detailed treatment as well as a discussion of M\o ller's admissibility conditions see \cite{13}.

\section{Hyperbolic observer}

Let P be uniformly accelerated along the $z$-axis with a world line
\beq
\label{eq:10}
\bar t =\frac{1}{g}\sinh g \tau, \quad \bar x=0, \quad \bar y=0, \quad \bar z= \frac{1}{g} (-1+\cosh g \tau).
\end{equation}
In this case (\ref{eq:1}) and (\ref{eq:2}) reduce to
\begin{eqnarray}
\label{eq:12}
t- \frac{1}{g} \sinh g (\eta -\rho)&= \sqrt{x^2+y^2+\left[z+\frac{1}{g}-\frac1{g}\cosh g (\eta -\rho)\right]^2}, \\
\label{eq:13}
t- \frac{1}{g} \sinh g (\eta+ \rho)&=- \sqrt{x^2+y^2+\left[z+\frac{1}{g}-\frac1{g}\cosh g (\eta +\rho)\right]^2}.
\end{eqnarray}
One can show on the basis of these equations that $(t,x,y,z) \to (\eta , \rho , \theta , \phi)$
is given by
\begin{eqnarray}
\label{eq:14} t&=&\frac{1}{g}(\cosh g \rho + \cos \theta \sinh g \rho)\sinh g \eta ,  \\
\label{eq:15} x&=&\left(\frac1g \sinh g \rho \right)\sin \theta \cos \phi , \qquad  y=\left(\frac1g \sinh g \rho \right)\sin \theta \sin \phi , \\
\label{eq:16} z&=&-\frac1g + \frac1g \left(\cosh g \rho + \cos \theta \sinh g \rho \right)\cosh g \eta  .
\end{eqnarray}
We note that as $g\to 0$, (\ref{eq:14})-(\ref{eq:16})
reduce to the standard spherical polar coordinates, as expected.
By treating $\rho$ to first order in (\ref{eq:14})-(\ref{eq:16}), it is possible to show that $(\rho , \theta , \phi)$
is indeed the spherical polar coordinate system constructed about P in an infinitesimal spatial neighborhood in the natural Fermi - Walker transported triad along the hyperbolic observer's world line. To this end, we recall that the Fermi coordinates are given in this (Rindler) space by \cite{10}
\begin{eqnarray}
\label{eq:16bis}
& t=\left(\frac{1}{g}+Z\right)\sinh g T, \quad x=X, \quad y=Y,\nonumber \\
& z=-\frac{1}{g}+\left(\frac{1}{g}+Z\right)\cosh g T.
\end{eqnarray}
Let us introduce polar coordinates via $X=\rho \sin \theta \cos \phi$, $Y=\rho \sin \theta \sin \phi$ and $Z=\rho \cos \theta$, where $\rho$ is treated to first order in comparison with the acceleration length $1/g$. Then, with  $T=\eta$, (\ref{eq:16bis}) and (\ref{eq:14}) - (\ref{eq:16}) are identical 
since  
$\cosh g \rho \simeq 1$ and $\sinh g \rho \simeq g \rho$ for $\rho \ll 1/g$.

The spacetime metric in radar coordinates is given by
\begin{eqnarray}
\label{eq:17}
\fl \rmd s^2 &= -(\cosh g\rho + \cos \theta \sinh g \rho)^2 \rmd \eta^2 +(\cosh 2g \rho + \cos \theta \sinh 2g \rho)\rmd \rho^2 \nonumber \\
\fl \quad &+\left(\frac1g \sinh g\rho\right)^2(-2g\sin \theta \rmd \rho \rmd \theta + \rmd \theta^2+ \sin^2 \theta \rmd \phi^2).
\end{eqnarray} 
It can be shown on the basis of the standard admissibility conditions given in appendix B that these radar coordinates are admissible if and only if 
\beq\label{eq:3.9}
     - g_{\eta \eta} = g_{\rho \rho} - g_{\rho \theta}^2 / g_{\theta \theta} > 0.        
\end{equation}
We note that $-g_{\eta \eta}=(z+\frac1g)^2-t^2=\left(Z+\frac1g\right)^2 > 0$ and vanishes at the Rindler horizon.

It  is possible to work out the components of $\xi^\mu=x^\mu-\bar x^\mu (\eta )$ in this case using equations (\ref{eq:10}) and (\ref{eq:14}) - 
(\ref{eq:16}). With respect to the natural tetrad frame of the observer P,
\begin{eqnarray}
&\lambda^\mu{}_{(0)}=(\cosh g \eta, 0,0, \sinh g\eta), \quad &\lambda^\mu{}_{(1)}=(0, 1,0, 0), \nonumber \\
&\lambda^\mu{}_{(2)}=(0, 0,1, 0), \qquad\qquad\qquad &\lambda^\mu{}_{(3)}=(\sinh g\eta, 0,0, \cosh g \eta),
\end{eqnarray}
the components of $\xi$ are $\xi_{(\alpha )}=\xi_\mu \lambda^\mu{}_{(\alpha )}$, so that in radar coordinates
\begin{eqnarray}
&\xi_{(0)}=0, \quad\qquad\qquad\qquad\qquad\qquad  &\xi_{(1)}=\left(\frac1g \sinh g\rho \right) \sin\theta \cos \phi, \nonumber \\
&\xi_{(2)}=\left(\frac1g \sinh g\rho \right) \sin\theta \sin \phi, \qquad & \xi_{(3)}=\frac1g (-1+\cosh g \rho)+ 
\left(\frac1g \sinh g\rho \right) \cos\theta .
\end{eqnarray}
Thus to lowest order in $g\rho \ll 1$,
\beq
\xi_{(1)}\simeq \rho \sin\theta \cos \phi, \quad
\xi_{(2)}\simeq\rho \sin\theta \sin \phi, \quad  \xi_{(3)}\simeq\rho \cos\theta ,
\end{equation}
as expected. The hypersurfaces of simultaneity are hyperplanes orthogonal to the world line of the hyperbolic observer, just as in the case of Fermi coordinates.

The construction of radar coordinates makes it evident that these are restricted to the Rindler wedge. As an example, consider an observer Q that moves along the $z$-axis:
$x=y=0$ and $z=v_0t$. For $t \ge 0$, Q starts at P and moves along the $\theta = 0$ direction according to 
\beq
\label{eq:18}
\cosh g \eta - v_0 \sinh g \eta=e^{-g \rho}
\end{equation}
from $\eta=0$ to $\eta=\eta_0$ given by
\beq
\label{eq:19}
\eta_0=\frac{1}g \ln \left( \frac{1+v_0}{1-v_0}\right),
\end{equation}
when it returns back to P. We note that $\dot \rho (\eta=0)=v_0$ and $\dot \rho (\eta=\eta_0)=-v_0$, while
$\ddot \rho (\eta=0\, {\rm or}\, \eta_0)=-g (1-v_0^2)$. In contrast, the corresponding result for Fermi coordinates is $-g ( 1 - 2 v_0^2 )$; see \cite{11}.
Here an overdot denotes differentiation with respect to the radar time $\eta$. For $\eta >\eta_0$, the motion is along $\theta=\pi$, hence
\beq
\label{eq:20}
\cosh g \eta - v_0 \sinh g \eta=e^{g \rho},
\end{equation}
so that as $\eta \to \infty$, Q approaches the Rindler horizon at $\rho=\infty$, since $- g_{\eta\eta} = e^{- 2 g \rho}$ in this case. In the limit that Q is a null ray with $v_0=1$, the motion is simply given by
$\rho=\eta$.
These results should be compared and contrasted with the results of \cite{11} based on Fermi coordinates.

\section{Uniformly rotating observer}

Consider next the world line of a uniformly rotating observer P parametrized by its proper time 
\beq
\label{eq:21}
\bar t=\gamma \tau, \quad \bar x=R\cos \Omega \tau, \quad \bar y=R\sin \Omega \tau, \quad \bar z=0.
\end{equation}
The observer moves with speed $v = R \Omega_0$ on a circle of radius $R$ with its center at the spatial origin of the global inertial coordinates. 
The azimuthal angle of P with respect to inertial coordinates is given by $\Omega_0 \bar t= \Omega \tau$, where $\Omega = \gamma \Omega_0$ and $\gamma$ is the Lorentz factor of P ,
\beq
                 \gamma = \sqrt{ 1 + R^2 \Omega^2 }.                  
\end{equation}
The radar coordinate system around P can be developed as in figure 1 with 
$\tau_1=\eta -\rho$ and $\tau_2=\eta+\rho$.
The following conditions must be satisfied:
\begin{eqnarray}
\label{eq:22}
\fl\quad
&& t-\gamma (\eta-\rho)=\sqrt{[x-R\cos \Omega (\eta-\rho)]^2+
[y-R\sin \Omega (\eta-\rho)]^2+z^2}:=\mathcal{R}_1, \\
\label{eq:23} \fl\quad
&& t-\gamma (\eta+\rho)=-\sqrt{[x-R\cos \Omega (\eta+\rho)]^2+[y-R\sin \Omega (\eta+\rho)]^2+z^2}:=-\mathcal{R}_2.
\end{eqnarray}
To simplify the analysis, consider the rotation $(x,y,z)\to (x',y',z')$
\beq
\label{eq:45n}
x=x'\cos\Omega \eta   -y'\sin \Omega \eta , \quad y=x'\sin\Omega \eta +y'\cos \Omega \eta , \quad 
z=z'.
\end{equation}
It follows that
\begin{eqnarray}
\label{eq:3-4}
\mathcal{R}_1&=& \sqrt{(x'-R\cos \Omega \rho)^2+
(y'+R\sin \Omega\rho)^2+z'{}^2}, \\
\label{eq:3-4a}
\mathcal{R}_2&=& \sqrt{(x'-R\cos \Omega \rho)^2+
(y'-R\sin \Omega\rho)^2+z'{}^2} .
\end{eqnarray}
From (\ref{eq:22})-(\ref{eq:23}) and (\ref{eq:3-4})-(\ref{eq:3-4a}) we get
\begin{eqnarray}
\label{eq:27}
&&(x'-R\cos \Omega \rho)^2+(y'+R\sin \Omega\rho)^2+z'{}^2= [t-\gamma(\eta-\rho) ]^2 ,  \\
\label{eq:28}
&&(x'-R\cos \Omega \rho)^2+(y'-R\sin \Omega\rho)^2+z'{}^2= [t-\gamma (\eta+\rho) ]^2 .
\end{eqnarray}
Adding and subtracting (\ref{eq:27}) and (\ref{eq:28}) result in
\begin{eqnarray}
\label{eq:29}
&&(x'-R\cos \Omega \rho)^2+y'{}^2+R^2 \sin^2 \Omega \rho + z'{}^2=\gamma^2\rho^2 +(t-\gamma\eta)^2 , \\
\label{eq:30} &&y'R\sin \Omega\rho= \gamma \rho (t-\gamma \eta ).
\end{eqnarray}
It proves useful to define the quantities $U$ and $V$,
\beq
\label{eq:31}
U= \frac{t-\gamma\eta}{R\sin\Omega \rho}, \quad V= \sqrt{(\gamma\rho)^2 -R^2\sin^2 \Omega \rho},
\end{equation}
where $(\gamma\rho)^2 -R^2\sin^2 \Omega \rho\ge 0$ and $V=0$ only when $\rho=0$. Based on these definitions, we have
\beq
\label{eq:32}
t= \gamma \eta +RU\sin \Omega \rho, \quad y'=\gamma \rho U
\end{equation}
and from (\ref{eq:28})
\beq
\label{eq:33}
(x'-R\cos \Omega \rho)^2+z'{}^2=(1-U^2)V^2,
\end{equation}
or
\beq
\label{eq:34}
\frac{(x'-R\cos \Omega \rho)^2}{V^2}+\frac{z'{}^2}{V^2}+U^2=1 .
\end{equation}
We now define the spherical polar angles $(\theta, \phi)$ as follows:
\beq
\label{eq:35}
\cos\theta = \frac{z'}{V},\quad  \sin\theta \cos \phi = \frac{(x'-R\cos \Omega \rho)}{V}, \quad \sin\theta \sin \phi=U,
\end{equation}
since it follows from equation (\ref{eq:34}) that  $z'{}^2/V^2\le 1$ and $U^2 \le 1$.
Hence
\begin{eqnarray}
\label{eq:36}
x'&=& R\cos \Omega \rho +V \sin \theta \cos \phi, \nonumber \\
y'&=& \gamma\rho \sin \theta \sin \phi, \nonumber \\
z'&=& V\cos \theta,
\end{eqnarray}
and from (\ref{eq:45n}), (\ref{eq:32}) and (\ref{eq:35})
\begin{eqnarray}
\label{eq:37}
t&=&\gamma\eta + R\sin \theta \sin \phi \sin \Omega \rho, \nonumber \\
x&=& (R\cos \Omega \rho +V \sin \theta \cos \phi)\cos \Omega \eta -(\gamma\rho \sin \theta \sin \phi)\sin \Omega \eta , \nonumber \\
y&=& (R\cos \Omega \rho +V \sin \theta \cos \phi)\sin \Omega \eta +(\gamma\rho \sin \theta \sin \phi)\cos \Omega \eta , \nonumber \\
z&=& V\cos \theta.
\end{eqnarray}
Note that for $\Omega = 0$ these relations reduce
to the 
usual spherical coordinates with origin at $(R,0,0)$:
\beq
\label{eq:39}
t=\eta, \quad x=R+\rho\sin\theta\cos\phi, \quad y=\rho\sin\theta\sin\phi, \quad z=\rho \cos \theta.
\end{equation}
To examine the physical significance of radar coordinates (\ref{eq:37}), it is important to consider the limiting case of $R=0$. The observer P is then fixed at the spatial origin of the global inertial frame and (\ref{eq:37}) reduces to $t=\eta$ and 
\beq\label{eq:38new}
\fl\quad
x=\rho \sin \theta \cos (\phi+\Omega_0 \eta),\quad  
y=\rho \sin \theta \sin (\phi+\Omega_0 \eta),\quad  
z=\rho \cos \theta .  
\end{equation}
This is the standard transformation to rotating Cartesian coordinates, which means that though P is fixed it is still noninertial as it refers its observations to axes rotating uniformly about the $z-$direction with frequency $\Omega_0$.
Indeed, these rotating axes are the natural orthonormal triad of P, as illustrated in appendix A and the radar angles
$(\theta , \phi)$ are defined with respect to these rotating axes.

At a given radar time $\eta$, which identifies the event $\bar x^\mu (\eta)$ on the world line of the reference observer P, the corresponding simultaneity hypersurface can be determined by eliminating the three spatial variables $\rho$, $\theta$ and $\phi$ from the four equations given in display (45). In contrast to the hyperbolic case, the form of this hypersurface for the rotating observer is not simple; nevertheless, by expanding $\sin \Omega \rho$ and $\cos \Omega \rho$ in powers of $\Omega \rho$ one can show explicitly that the hypersurface perpendicularly intersects the world line of P.

Let us compute the components of the vector $\xi^\mu=x^\mu-\bar x{}^\mu (\eta)$ for the uniformly rotating observer P using (\ref{eq:21}) and 
(\ref{eq:37}). 
With respect to the natural tetrad frame (\ref{eq:A1}) given in appendix A, the measured components of $\xi$, $\xi_{(\alpha )}=\xi_\mu \lambda^\mu{}_{(\alpha )} $, are given by
\begin{eqnarray}
\xi_{(0)}&=& \gamma R (\Omega \rho -\sin \Omega \rho)\sin \theta \sin \phi ,\nonumber \\
\xi_{(1)}&=& R(-1+\cos \Omega \rho)+V\sin \theta \cos \phi ,\nonumber \\
\xi_{(2)}&=& (\gamma^2 \rho-R^2\Omega \sin \Omega \rho)\sin \theta \sin \phi ,\nonumber \\
\xi_{(3)}&=& V\cos \theta .\nonumber \\
\end{eqnarray}
It is important to note that this vector is not in general orthogonal to the world line of P, so that the simultaneity hypersurfaces in this case differ from the normal hyperplanes employed in the case of Fermi coordinates. For $\Omega \rho \ll 1$, $\cos \Omega \rho \simeq 1$, 
$\sin \Omega \rho \simeq \Omega \rho$ and $V\simeq \rho$; therefore, $\xi_{(0)}\simeq 0$ and 
\beq
\xi_{(1)}\simeq \rho \sin \theta \cos \phi, \quad
\xi_{(2)}\simeq \rho \sin \theta \sin \phi, \quad
\xi_{(3)}\simeq \rho  \cos \theta,
\end{equation}
as expected.

It is interesting to return to the problem of the radar distance from P to the center of the circle of radius $R$.
This center has inertial coordinates $(t, x=y=z=0)$; therefore (\ref{eq:37}) implies that
\beq\label{eq:39new}
\eta=t/\gamma, \, \rho=R/\gamma, \, \theta=\pi/2, \, \phi=\pi,
\end{equation}
in agreement with section 2. As is clear from figure 4 in appendix A, the center of the circle is always at $(\theta,\phi)=(\pi/2,\pi)$ for the reference observer P.

The metric in radar coordinates has a complicated form in this case; therefore, to simplify the analysis of the admissibility of radar coordinates we use the coordinates $(\eta, x', y', z')$ instead. Let us note that $x'$, $y'$ and $z'$ are independent of $\eta$ and they have their origin at the center of the circle, while $(\rho, \theta, \phi)$ has its origin at P. It turns out, as demonstrated in detail in appendix B, that a further coordinate transformation
to the actual radar coordinates, $(\eta , x', y', z')\to (\eta, \rho, \theta, \phi)$, does not affect the admissibility analysis  presented here.
We assume that $(x', y', z')\to (\rho, \theta, \phi) $ is a proper coordinate transformation and define
\beq
\label{star}
W(x', y', z')= R \sin \theta \sin \phi \sin \Omega \rho.
\end{equation}
Then, in terms of $(\eta, x', y', z')$ the metric is
\beq\label{2star}
\fl \rmd s^2 =-(\gamma \rmd \eta + \rmd W)^2 + \Omega^2 (x'{}^2+ y'{}^2)\rmd \eta^2+
2\Omega (x' \rmd y'-y' \rmd x')\rmd \eta +\rmd x'{}^2+\rmd y'{}^2+
\rmd z'{}^2.
\end{equation}
It follows from the general form of this metric that the conditions for the admissibility of
the $(\eta, x', y', z')$ coordinate system reduce to
\beq\label{3star}
-g_{\eta\eta}=\gamma^2-\Omega^2 (x'{}^2+ y'{}^2) >0.
\end{equation}
Thus the boundary of the admissible region for radar coordinates is given by $g_{\eta\eta}=0$.

It follows from (\ref{3star}) that
\beq
\label{eq:38}
\fl
g_{\eta\eta}= -(1+\Omega^2R^2)[1-(\Omega\rho)^2 \sin^2 \theta\sin^2 \phi]+\Omega^2(R\cos \Omega \rho+V\sin\theta\cos\phi)^2,
\end{equation}
so that the boundary surface $\rho=\rho(\theta, \phi)$ is given by
\beq
\label{eq:40}
\fl\quad
-(1+\Omega^2R^2)[1-(\Omega\rho)^2 \sin^2 \theta\sin^2 \phi]+\Omega^2(R\cos \Omega \rho+V\sin\theta\cos\phi)^2=0 .
\end{equation}
This surface has the topology of a cylinder about the observer's $Z-$axis (cf. appendix A), since for $\theta=0$ or  $\pi$ we have $-g_{\eta \eta} =1+R^2\Omega^2\sin^2 \Omega\rho$, which is always positive. Moreover, for $\rho \gg R / \gamma$, equation (\ref{eq:40}) implies that $\rho \sin \theta \sim 1 / \Omega$, so that the asymptotic radius of the cylinder is $\sim \Omega^{-1}$. 
For $\theta = \pi / 2$, 
the boundary surface reduces to a closed curve that is depicted in figure 3, where $\Omega \rho \sin \phi$ is plotted versus $\Omega \rho \cos \phi$ for $\Omega R=1$.

\begin{figure}[h] 
\typeout{*** EPS figure 3}
\begin{center}
\includegraphics[scale=0.40,angle=-90]{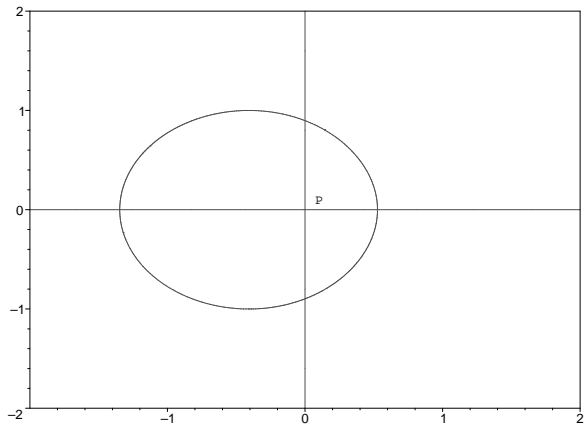} 
\end{center}
\caption{Schematic representation of the closed boundary curve $\rho=\rho(\pi/2 , \phi)$. We plot $\Omega \rho \sin \phi$  versus $\Omega \rho \cos \phi$ for the special case where $\Omega R=1$ and hence $v=1/\sqrt{2}$. 
}  
\label{fig:3}
\end{figure}

\section{Discussion }

The theoretical construction of an extended frame of reference is a basic problem in the theory of relativity and has been the subject of many recent investigations \cite{13,12,14}. The present paper has been devoted to the construction and investigation of radar coordinates. For the sake of simplicity , we have limited the scope of our main work to accelerated observers in Minkowski spacetime and have found that radar coordinates have limited utility in terms of the concept of radar distance. Moreover, the domain of applicability of radar coordinates is limited just as it is with Fermi coordinates. Once an observer is accelerated, the absolute character of its acceleration is reflected in the existence of local acceleration lengths. While these locally limit the pointwise applicability of the hypothesis of locality \cite{3}, they also determine the scale of the domain of validity of physically - motivated coordinate systems established around the world line of the observer.  Radar coordinates may be interesting and useful, especially in connection with the synchronization of distant clocks; however, they cannot in general replace the more basic Fermi coordinates. 

For a general observer in a gravitational field, the light signals follow null geodesics and the calculations involved in the determination of radar coordinates would then become considerably more complicated. Moreover, in addition to possible acceleration scales, curvature radii must be taken into consideration. For distances beyond the curvature radii, the uniqueness of null geodesics or the spacelike geodesics used to define the angular radar coordinates is not assured.

The general issues discussed in this paper become relevant whenever electromagnetic or other (null) signals are used for spacetime determinations.
In this connection, we must mention the GPS system, where the coordinates of an event are determined, in principle, by the simultaneous reception of four GPS signals \cite{ash, rov, bla}. We recall that the radar coordinate system has the important property that it essentially reduces to
the Fermi system when the range $\rho$ is sufficiently small compared to the characteristic lengthscale of the observer. Let us note that for observers on the Earth, $1/\Omega_\oplus\simeq  28\,$ AU and $1/g_\oplus \simeq 1\,$ lt-yr , so that for the GPS system the corresponding $\rho/L$ is generally very small.

The scheme of setting up coordinate systems about the world lines of observers may be called the 1+3 splitting of spacetime, since the time coordinate is essentially distinguished as it is the proper time of an observer. A complementary scheme is based on a natural foliation of spacetime by spacelike hypersurfaces. Admissible 3+1 splittings have been discussed in \cite{15}. Moreover, a dynamical realization of the 3+1 scheme in the case of weak gravitational fields is contained in \cite{16}.

The present knowledge of the structure of the universe is dependent to a large extent upon the reception of electromagnetic signals from distant astronomical bodies as well as communication with spacecraft via radio signals that are transponded back to the Earth. This circumstance provides the motivation to investigate generalizations of the radar coordinate system as, for instance, the approach recently proposed in \cite{13}.

\appendix

\section{Fermi coordinates for uniformly rotating observer}
Consider the uniformly rotating observer P  in section 4 with its world line given by (\ref{eq:21}).
The natural orthonormal tetrad of this observer is given in $(t,x,y,z)$ coordinates  by \cite{3}
\begin{eqnarray}\label{eq:A1}
\lambda_{(0)}^\mu &=& \gamma (1, -v \sin \varphi, v \cos \varphi, 0), \nonumber \\
\lambda_{(1)}^\mu &=& (0, \cos \varphi, \sin \varphi ,0), \nonumber \\
\lambda_{(2)}^\mu &=& \gamma (v,  -\sin \varphi ,\cos \varphi,0), \nonumber \\
\lambda_{(3)}^\mu &=& (0, 0, 0 ,1), 
\end{eqnarray}
where $v=R\Omega_0$ and $\varphi=\Omega_0 \bar t = \Omega \tau $. The corresponding Fermi normal coordinates
$(T,X,Y,Z)$ are given by \cite{3}
\begin{eqnarray}\label{eq:A2}
T&=& \frac{1}{\gamma}(t-\gamma v Y), \nonumber \\
X&=&-R+x \cos \Omega T+ y \sin \Omega T, \nonumber \\
Y&=& \frac{1}{\gamma}(-x \sin \Omega T + y \cos \Omega T), \nonumber \\
Z&=& z.
\end{eqnarray}


\begin{figure}[ht]
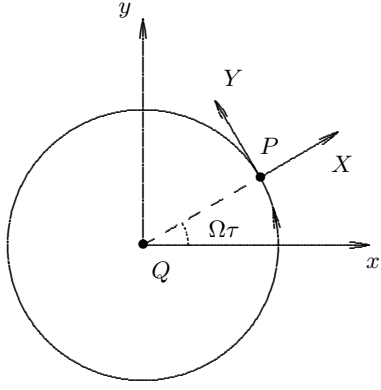

$$ \vbox{
\beginpicture
  \setcoordinatesystem units <0.6cm,0.6cm> point at 0 0 
\circulararc 360 degrees from 3 -3 center at 0 -3

\setdashes <5pt>
  \plot  0 -3   2.6 -1.5    / 
\setsolid 
\setdashes <1pt>
\circulararc 30 degrees from 1 -3 center at 0 -3 
\setsolid

\put {\mathput{\bullet}}            at 2.6 -1.5 
\put {\mathput{\bullet}}            at 0 -3
\put {\mathput{Q}}         at 0.4 -3.6
\put {\mathput{P}}         at 2.8 -0.8
\put {\mathput{x}}         at 5.1 -3.4
\put {\mathput{y}}         at -0.4 2.2
\put {\mathput{Y }}        at  2  0.7
\put {\mathput{X} }        at  4.5  -1.2
\put {\mathput{\Omega \tau}}               at  1.8  -2.65
\arrow <.25cm> [.2,.3]    from  2.6 -1.5  to 1.6 0.2
\arrow <.25cm> [.2,.3]    from  2.6 -1.5  to 4.31 -0.5
\arrow <.25cm> [.2,.3]    from  0 -3  to 0 2
\arrow <.25cm> [.2,.3]    from  0 -3  to 5 -3
\arrow <.25cm> [.2,.3]    from  2.95 -2.5  to 2.89 -2.2
\endpicture}$$
\vspace{2cm}
\caption{Local axes of the rotating observer P.}
\end{figure}

The center of the circle is represented by Q: $(t,x=y=z=0)$ with Fermi coordinates $T=t/\gamma$, $X=-R$ and $Y=Z=0$  as in figure 4. Thus the Fermi distance between P and Q is the radius of the circle $R$, as expected.
Transforming $(X,Y,Z)$ to spherical polar coordinates $(\hat \rho, \theta, \phi)$, we note that Q is at 
$\hat \rho=R$, $\theta=\pi/2$ and $\phi=\pi$. Moreover, with $\hat \rho=\rho$ very small and $T=\eta$, (\ref{eq:A2}) is identical with (\ref{eq:37}) when $\Omega\rho$ is treated to first order (i.e. $\cos \Omega \rho \simeq 1$, $\sin\Omega \rho\simeq \Omega \rho$ and $V\simeq \rho$); as discussed above, this is simply a consequence of the fact that the spherical polar coordinates are defined with respect to the local spatial frame of the observer P.
The spacetime metric in Fermi coordinates is
\begin{eqnarray}
\fl\quad
\rmd s^2 &= - \gamma^2 [1-\Omega_0^2(X+R)^2-\gamma^2 \Omega_0^2 Y^2]\rmd T^2 + 2 \gamma^2 \Omega_0 (X\rmd Y- Y \rmd X)\rmd T\nonumber \\
\fl\quad
&+ \rmd X^2+\rmd Y^2 +\rmd Z^2.
\end{eqnarray}
The Fermi coordinates are admissible for $\Omega_0^2 (X+R)^2+\gamma^2 \Omega_0^2 Y^2 <1.$
The boundary region is an elliptic cylinder whose axis coincides with the $Z-$axis  and is characterized in the $(X,Y)$-plane by an ellipse with its center at Q. The ellipse has semimajor axis $1/\Omega_0$ and semiminor axis
$1/(\gamma \Omega_0)$; in fact, this ellipse with eccentricity $v<1$ can be considered in terms of a circle of radius $1/\Omega_0$ that is Lorentz-Fitzgerald contracted along the direction of motion of P, which occupies a focus of this ellipse.

\section{Standard admissibility conditions}

Consider a region of spacetime where events have been assigned coordinates $x^\mu = (t, x^i)$ such that the spacetime interval takes the form
$\rmd s^2=g_{\mu\nu}\rmd x^\mu \rmd x^\nu$ with signature $+2$. The fundamental observers in this region are the observers at rest, i.e. $x^i=$ constant for each observer. The standard admissibility conditions for this system of coordinates refer to the local measurements of the fundamental observers using clocks and infinitesimal measuring rods. The proper time $\tau$ of each fundamental observer is given by $-\rmd \tau^2=g_{00}\rmd t^2$, so that $g_{00}<0$
is the first admissibility condition. Let
\beq
u^\mu=(\frac{1}{\sqrt{-g_{00}}},0,0,0)
\end{equation} 
be the four-velocity field of the fundamental observers.
Each has access to infinitesimal measuring sticks that exist in an infinitesimal hypersurface orthogonal to  its world line, i.e. $g_{\mu\nu}u^\mu \rmd x^\nu =0$. The infinitesimal orthogonal hypersurface is thus given by $g_{00}\rmd t + g_{0i}\rmd x^i=0$, which combined with 
\beq\label{B2}
g_{\mu\nu}\rmd x^\mu \rmd x^\nu= \frac{1}{g_{00}}(g_{00}\rmd t + g_{0i}\rmd x^i)^2+\gamma_{ij}\rmd x^i \rmd x^j\, ,
\end{equation}
where
\beq
\gamma_{ij}=g_{ij}-\frac{g_{0i}g_{0j}}{g_{00}},
\end{equation}
implies that the length of an infinitesimal measuring rod is $\rmd l^2=\gamma_{ij}\rmd x^i \rmd x^j$. It follows that other standard admissibility conditions are required to ensure that $\rmd l^2$ is positive. A theorem of linear algebra \cite{hof} states that for a symmetric $n\times n$ matrix $M$,
$M_{ij}\rmd x^i \rmd x^j$ is positive if and only if the principal minors of $M$ are all positive, i.e.
\beq
\label{B4}
{\rm det}\, 
\pmatrix{M_{11}&.\quad .\quad . &M_{1k}\cr
	. & &.\cr
	. & &.\cr
	. & &.\cr
	M_{k1}&.\quad .\quad . &M_{kk}\cr}
>0, \quad {\rm for}\quad k=1, \ldots , n .
\end{equation}
Moreover, it can be shown that  $M$ is positive if and only if there exists an invertible matrix $N$ such that $M=N^\dag N$, where $N^\dag$
is the transpose of $N$. Then, $M_{ij}\rmd x^i \rmd x^j$ can be expressed in matrix notation as $(N \rmd x)^\dag(N \rmd x)$, which is manifestly positive; that is, if $\rmd x\not =0$, then $N \rmd x \not =0$, since $N$ is invertible, and so $M_{ij}\rmd x^i \rmd x^j >0$ \cite{hof}.

The admissibility conditions (\ref{B4}) are the same as those given in \cite{5} on the basis of a different analysis involving the determination of radar distance between two infinitesimally close observers. Let us now apply these conditions to the spacetime metric in radar coordinates for the case of 
the rotating observer in section 4. The transformation from the inertial coordinates to radar coordinates in (\ref{eq:37}) may be considered to be a combination of $(t,x,y,z) \to (\eta, x', y', z')$ given by $t=\gamma \eta +W(x',y', z')$ and (\ref{eq:45n}) together with 
$(\eta, x', y', z')\to (\eta , \rho , \theta , \phi)$ given by (\ref{eq:36}).
Consider first the transformation $(t,x,y,z) \to (\eta, x', y', z')$ that results in the metric (\ref{2star}). It is straightforward to show that for this metric $g_{\eta\eta}<0$ implies that the relations (\ref{B4}) are all satisfied.
This means that for $g_{\eta\eta}<0$, there exists an invertible matrix $S$ such that
$\gamma'{}_{ij}\rmd x'{}^i \rmd x'{}^j=(S\rmd x')^\dag (S\rmd x')$. Next, we apply the transformation (\ref{eq:36}) to the metric (\ref{2star}) to get the general metric in radar coordinates, which are admissible if $\gamma_{ij}\rmd x^i \rmd x^j$ is positive.
Let $H$ be the Jacobian matrix of the transformation (\ref{eq:36}).
This matrix is invertible if the corresponding Jacobian, $J={\rm det}\, H$,  is nonzero. It is possible to express $J$ as 
\beq
J=\gamma \rho^2 \sin \theta F,
\end{equation} 
where $F(\rho, \theta, \phi)$ is given by
\beq
\label{B6}
F=\frac{V}{\rho}\left[\frac{\rmd V}{\rmd \rho}(\cos^2 \theta + \sin^2 \theta \cos^2 \phi)+\frac{V}{\rho}\sin^2\theta \sin^2\phi-R\Omega 
\sin \theta \cos \phi \sin \Omega \rho\right] .
\end{equation}
It is simple to see from (\ref{B6}) that $F$ is always positive along the axis of rotation $(\theta =0, \pi)$, $F\to 1$ for $\rho \to 0$ and $F> 0$ for 
$\rho \to \infty$. Moreover, $F(\rho, \theta, \pi/2)\ge 1$. Extensive numerical analysis has shown that $F$ is indeed positive.
 On this basis, we assume that $H$ is invertible. Then it follows from the structure of (\ref{B2}) that in radar coordinates 
$\gamma_{ij}\rmd x^i \rmd x^j=(SH\rmd x)^\dag(SH\rmd x)$, where $SH$ is an invertible matrix.
Thus $(\gamma_{ij})$ is a positive matrix and  $g_{\eta\eta}<0$ is the only condition that we need in order to ensure that the radar coordinates for the rotating observer are admissible.

\end{document}